\pdfoutput=1

\documentclass[%
 reprint,
 superscriptaddress,
 nobibnotes,
 amsmath,amssymb,
 aps,
 showkeys,
 longbibliography,
]{revtex4-2}

\usepackage{graphicx}
\usepackage{dcolumn}
\usepackage{bm}
\usepackage{chemformula}
\usepackage[separate-uncertainty=true,separate-uncertainty,multi-part-units=single]{siunitx}
\usepackage{hyperref}
\hypersetup{colorlinks=true,allcolors=magenta}

\begin{document}

\title{Adapted Thermodynamical Model for the Prediction of Adsorption in Nanoporous Materials}

\author{F. Stavarache}
    \affiliation{Materials Simulation \& Modelling, Department of Applied Physics, Eindhoven University of Technology, 5600 MB, Eindhoven, The Netherlands}
\author{A. Luna-Triguero}
    \affiliation{Energy Technology, Department of Mechanical Engineering, Eindhoven University of Technology, 5600 MB, Eindhoven, The Netherlands}
    \affiliation{Eindhoven Institute for Renewable Energy Systems (EIRES), Eindhoven University of Technology, Eindhoven 5600 MB, The Netherlands}
\author{S. Calero}
    \affiliation{Materials Simulation \& Modelling, Department of Applied Physics, Eindhoven University of Technology, 5600 MB, Eindhoven, The Netherlands}
    \affiliation{Eindhoven Institute for Renewable Energy Systems (EIRES), Eindhoven University of Technology, Eindhoven 5600 MB, The Netherlands}    
\author{J. M. Vicent-Luna}
    \email[Corresponding author: ]{j.vicent.luna@tue.nl}
    \affiliation{Materials Simulation \& Modelling, Department of Applied Physics, Eindhoven University of Technology, 5600 MB, Eindhoven, The Netherlands}
    \affiliation{Eindhoven Institute for Renewable Energy Systems (EIRES), Eindhoven University of Technology, Eindhoven 5600 MB, The Netherlands}

\date{\today}

\begin{abstract}

In this paper, we introduce a novel, adapted approach for computing gas adsorption properties in porous materials. We analyze the Dubinin-Polanyi's adsorption model and investigate various frameworks to estimate its required essential components. Those are linked to physicochemical properties of the adsorbates, such as the vapor saturation pressure and density in the adsorbed state. To conduct this analysis, we obtain adsorption isotherms for several metal-organic frameworks, encompassing a range of pore sizes, shapes, and chemical compositions. We then apply and evaluate multiple combinations of models for saturation pressure and density.

After the evaluation of the method, we propose a working thermodynamic model for computing adsorption isotherms, which entails using the critical isochore as an approximation of the saturation pressure above the critical point and applying Hauer's method with a universal thermal expansion coefficient for density in the adsorbed state. This framework is applicable not only to simulated isotherms but also to experimental data from the literature for various molecules and structures, demonstrating robust predictive capabilities and high transferability. Our method showcases superior performance in terms of accuracy, generalizability, and simplicity compared to existing methods currently in use. For the first time, a method starting from a single adsorption curve and based on physically interpretable parameters can predict adsorption properties across a wide range of operating conditions.

\end{abstract}


\maketitle


\section{Introduction}
\label{sec:intro}
Gas adsorption in nanoporous materials has emerged as an efficient and environmentally friendly technology, providing a versatile alternative to conventional gas storage and separation methods.\cite{Ads01,Ads02,Ads03,Ads04} It plays a pivotal role in our society, finding applications across various fields, including carbon dioxide capture, environmental protection, energy storage, healthcare, air quality improvement, and circular economy practices.\cite{App01,App02,App03,App04} This technology holds significant promise in addressing the global environmental and sustainability challenges of our time. Materials such as zeolites, carbon-based materials, polymers, and metal-organic frameworks, among others, offer diverse chemical compositions and structural architectures suitable for gas adsorption and storage.\cite{porous01,porous02,porous03} The range of porous materials continues to expand with ongoing advances in material synthesis. However, most of these materials remain limited to laboratory settings due to the absence of predictive tools for assessing their performance in specific applications. As a result, researchers face the challenge of identifying optimal materials from an extensive array of choices. Therefore, accurately predicting adsorption behavior under varying temperature and pressure conditions is essential for describing the entire gas adsorption process. 

In the past century, extensive research on adsorption paved the way for industrial applications of adsorption-related technologies. Examples include pressure swing adsorption (PSA),\cite{Broom2013,PSA01,PSA02,PSA03} where system pressure is adjusted to capture or release molecules. Additionally, temperature swing adsorption (TSA),\cite{TSA01,TSA02} and pressure-temperature swing adsorption (PTSA)\cite{PTSA01,PTSA02} provide greater control over gas filling and discharging, making them valuable for applications like gas separation, purification, and even carbon dioxide removal from the atmosphere. The operating principle of these techniques is to modify the working conditions to control the adsorption amount. Obtaining adsorption equilibrium data under various conditions is a challenging task. Existing literature typically offers single adsorption isotherms for specific adsorbent-adsorbate pairs at certain temperatures, limiting the establishment of relationships between loading, pressure, and temperature. Material characterization becomes challenging due to the dependence on molecular-surface interactions, and there is a growing demand for determining adsorption properties under diverse environmental conditions. This not only sheds light on potential material usage but also deepens our understanding of the physical interactions occurring within them \cite{Broom2013}.

In this context, various mathematical models have been proposed to describe adsorption processes in different scenarios, categorized into different types, such as kinetics, thermodynamics, and, in the context of this work, the adsorption potential theory \cite{Foo2010}. However, most of these models are primarily for understanding and describing adsorption processes and lack predictive power since they primarily focus on fitting adsorption isotherms to equations. An exception is the adsorption potential theory, which establishes that all adsorption isotherms for a specific adsorbate-adsorbent pair can be transformed into the same characteristic curve \cite{Dubinin1960}. This theory suggests that a single adsorption isotherm can determine the adsorption behavior across a wide range of temperature and pressure conditions, making it a powerful predictive theory.

In this paper, we investigate the thermodynamic theory of adsorption potential proposed by Dubinin and Polanyi, which assumes pore filling as the dominant mechanism in micropores. While this hypothesis has opened debates among scientists due to its limitations in describing the layer-by-layer adsorption mechanism, we demonstrate its predictive power and efficiency in determining gas adsorption properties in diverse porous materials. In contrast to traditional models that rely on numerous adjustable parameters, this approach stands out for its minimal use of such parameters, enhancing its predictive capabilities. This innovation has the potential to significantly impact the field of adsorption by providing a more accurate and efficient way to characterize gas adsorption behavior.

In recent years, numerous experiments and studies have been conducted to validate the potential theory, with most producing promising but limited results. Wang et al. \cite{Wang2010} demonstrated that predictions for xenon adsorption in activated charcoal at 303 K and 323 K, based on an adsorption isotherm at 293 K, aligned well with experimental data for those respective temperatures. Du and Wu \cite{Du2007} developed a formula for the adsorption function of molecular hydrogen on zeolites using the adsorption potential theory. Hong et al. \cite{Yang2010} used the characteristic curve of molecular nitrogen on coal at 77 K to predict the adsorption isotherm at 303 K, achieving a match with experimental data. However, many of these studies have predominantly focused on carbon-based adsorbents like coal, carbon molecular sieves, and activated carbons \cite{Song2018}, often with a limited range of adsorbates. Furthermore, adsorption theory requires mathematical modeling of both the saturation pressure of a gas and its density within small cavities. Many of the models proposed for this purpose either lack precision or rely on existing experimental data and involve fitted parameters, making them less adaptable from one adsorbate to another and casting doubt on their predictive nature. Consequently, the applicability of this theory to other complex structures, such as zeolites, metal-organic frameworks (MOFs), covalent-organic frameworks, or diverse adsorbates, remains unclear. 

Our study focuses on using the adsorption potential theory as a predictive model for adsorption properties across various nanoporous structures and diverse adsorbates. In the initial phase, we employ Grand Canonical Monte Carlo (GCMC) simulations to compute the adsorption properties of carbon dioxide, methane, and nitrogen within several MOFs, specifically Co-MOF-74, IRMOF-1, MIL-47, MOF-1, and ZJU-198. These MOFs cover a wide range of pore sizes and shapes, and chemical compositions, ensuring that different mechanisms are involved in the adsorption behavior. In the subsequent step, we expand our investigation to include experimental data retrieved from the literature. This dataset incorporates adsorption data for argon, propane, methanol, and ammonia within five additional MOFs (MOF-467, Cu-INAIP, Co$_3$-HCOO$_6$, STAM-1, and Cu-BTC), as well as a zeolite (LTA4A). We examine existing methods, beginning with those proposed by Song et al. \cite{Song2018}, and extend the list to encompass novel options that were previously unexplored. These methods are then applied to predict isotherms across numerous porous materials, which are subsequently compared to simulated results. We present a comprehensive performance comparison of all methods, with the most effective one subsequently validated through experimental data from the literature, corroborating the predictive capability of our proposed approach.

\section{Methodology}
\label{sec:methods}
\subsection{Theory}

\subsubsection{The Polanyi adsorption potential theory}

According to the adsorption potential theory, molecules are attracted to surfaces via a potential that is uniquely described by the molecule's proximity to the surface, thus unaffected by the neighboring molecules inside the potential \cite{Polanyi1963}. Moreover, according to the theory, condensation of the molecules happens once the potential they are subject to equals or exceeds the work needed to bring the vapor particles to liquid state \cite{Chiou2003}. This differs from Langmuir adsorption, where condensation is reached once a mono-layer has formed over the adsorption surface \cite{Chiou2003}.

The equation for the potential energy required to condense vapor particles from environmental conditions, also called adsorption potential, is given by \cite{Manes1969}:

\begin{equation} \label{eq:potential}
    A = RT \ln{\left( \frac{P_{sat}}{P} \right)}
\end{equation}

\noindent where $R$ is the universal gas constant, $T$ is the temperature of the adsorbate, $P_{sat}$ is the saturation pressure, and $P$ is the pressure of the adsorbate. 

Since the potential energy felt by the particles attracted to a solid surface is position dependent, any potential energy value is represented by a surface area called the equipotential surface. Particles found on the same equipotential surface are subject to the same attraction energy to the solid surface. For any pair of environmental conditions, there is a surface area that corresponds to the potential energy required to condense the vapor, where the particles below this surface will condense while the particles above will remain in vapor form. As such, each value of the potential energy required to condense the vapor should correspond to a unique amount of particles condensed on the surface.

When applying potential theory on nanoporous materials that have complex structures, it results that for any filling ratios of the material, there exists a unique adsorption potential. To this date, the most accurate way of describing this filling volume was proposed by Dubinin \cite{Chiou2003} and is given by the following formula:

\begin{equation} \label{eq:volume}
    W = \frac{q}{\rho_{ads}}
\end{equation}

\noindent where $q$ is the adsorbed amount, and $\rho_{ads}$ is the density of the adsorbate within the adsorbent. By plotting the adsorption potential against the filling volume a characteristic line emerges, called the adsorption characteristic curve, which is specific for each adsorbate-adsorbent pair. Moreover, since the adsorption potential is unique for each filling volume, temperature and pressure changes do not affect the shape or position of the characteristic curve. Thus, the potential theory implies that it is possible to predict the behavior of adsorption for any range of pressure-temperature conditions if the characteristic curve of the given adsorbate-adsorbent pair is known in advance.

\subsubsection{Adsorbate saturation pressure}

In a closed system, transitions between different states of a solution occur depending on the system conditions, two such transitions being evaporation and condensation. Even though under environmental conditions, both transitions happen at the same time, system conditions usually favor only one of the transitions, with the solution tending towards a homogeneous phase state. However, for each temperature, there is a specific pressure at which the rate of condensation and evaporation are equal, called saturation pressure. At saturation conditions, the liquid and vapor states of the same substance can coexist. To obtain accurate values for the adsorption potential, a good estimation for the vapor-liquid saturation pressure of the adsorbate inside the nanoporous media is needed. However, the saturation pressure, also called the curve of coexistence, is only determined in the subcritical range, where there is a clear boundary between the vapor and liquid phases of the adsorbate. Surpassing the critical point results in supercritical behavior, where a substance exists in a mixed state. For this reason, adsorption above the critical pressure of an adsorbate requires estimations of the so-called virtual saturation pressure or curve of hidden boiling \cite{Gorbaty1998}.

Modeling of the saturation pressure above the critical point comes with many challenges, and all attempts can only be regarded as empirical, as in theory, there is no distinction between gasses and fluids above the critical point, let alone a defined boundary between the two \cite{Gorbaty1998}. Moreover, the models proposed should ideally be simple and general as to allow for the calculation of the adsorption potential for different adsorbates with little to no work. For this reason, attempts to differentiate between liquid-like and gas-like behavior in the supercritical phase that makes use of quantum mechanics are avoided in this work.

\begin{itemize}
    \item {Dubinin's method}

In 1960, while reviewing Polanyi's adsorption potential theory, Dubinin proposed the following method for estimating the saturation pressure $P_{sat}$ of the adsorbate \cite{Dubinin1960}:

\begin{equation} \label{eq:dubinin}
    P_{sat} = P_{C} \left( \frac{T}{T_{C}} \right) ^{2}
\end{equation}

\noindent where $P_{C}$ and $T_{C}$ are the critical pressure and temperature of the adsorbate and $T$ is its temperature. Dubinin successfully used this approximation to calculate the characteristic curves of xenon in two different activated carbons, together with carbon dioxide in silica gel.

\item{Amankwah's method}

Following up on Dubinin's work, Amankwah \cite{Amankwah1995} proposed in 1995 an adapted method for estimating the saturation pressure of the adsorbate in the supercritical range , given by:

\begin{equation} \label{eq:amankwah}
    P_{sat} = P_{C} \left( \frac{T}{T_{C}} \right) ^{k}
\end{equation}

\noindent where $k$ is a specific fit parameter for each adsorbate-adsorbent pair. The reason for that was the inaccurate results of Dubinin's approximation for some adsorbates such as hydrogen. For this method, experimental isotherms were used to determine the exponent $k$ for which there is full overlap between the characteristic curve. 

While it is possible to improve the fitting power of this method by computing $k$ parameters that are both temperature dependent and adsorbate-adsorbent specific \cite{Srinivasan2011}, the simplicity of Amankwah's equation makes it a more viable choice for most industrial applications. Unfortunately, to this date, there are very limited $k$ values available, which limits the use of this equation.

\item{Peng-Robinson equation of state}

Aiming to account for the shortcomings of the Soave-Redlich-Kwong equation of state when used for vapor-liquid equilibrium calculations, Peng and Robinson proposed in 1976 \cite{Peng1976} a new two-constant equation of state given by:

\begin{equation} \label{eq:peng-robinson}
    P = \frac{RT}{v-b} - \frac{a(T)}{v (v + b) + b (v - b)}
\end{equation}

where $P$ is the environmental pressure, $R$ is the universal gas constant, $v$ is the molar volume, $a$ a temperature dependent variable, and $b$ is a constant. The latter two values are defined as follows:

\begin{equation} \label{eq:aba}
    a(T) = 0.45724 \frac{R^{2} T_{C}^{2}}{P_{C}} \alpha(T)
\end{equation}

\begin{equation} \label{eq:abb}
   b = 0.07780 \frac{R T_{C}}{P_{C}}
\end{equation}

\noindent where $T_{C}$ and $P_{C}$ represent the critical temperature and pressure respectively. The scaling of the parameter $a$ at temperatures different than $T_{C}$ is given by the $\alpha$ function, which is defined as:

\begin{equation} \label{eq:alpha_preos}
    \alpha(T) = \left( 1 + \kappa \left( 1 - \sqrt{\frac{T}{T_{C}}} \right) \right)^{2}
\end{equation}

\begin{equation} \label{eq:kappa}
    \kappa = 0.37464 + 1.54226 \omega - 0.26992  \omega^{2}
\end{equation}

\noindent where $\omega$ is the acentric factor of the molecule.

While eq. \ref{eq:peng-robinson} does not provide additional information for vapor-liquid equilibrium calculations, its polynomial form does:

\begin{equation} \label{eq:peng-robinson-polynomial}
    \begin{split}
     Z^{3} - (1 - B) Z^{2} + (A - 3 B^{2} - 2 B) \cdot \\
    \cdot Z - (A B - B^{2} - B^{3}) = 0
     \end{split}
\end{equation}

Here, $Z$ is defined as the compressibility factor, with the smallest positive root corresponding to the compressibility factor of the liquid, and the largest one to that of vapor. Moreover, the variables $A$ and $B$ are defined in terms of the previously established values as such:

\begin{equation} \label{eq:AB}
    A = \frac{a P}{R^{2} T^{2}},\ \ \ 
    B = \frac{b P}{R  T}
\end{equation}

The advantage of this equation is that it provides a direct way of computing the compressibility factor of a gas, which can then be used to calculate its fugacity coefficient and, thus, the fugacity of the gas. Peng and Robinson proposed the following method for determining the fugacity coefficient $\phi$:

\begin{equation} \label{eq:fugacity_coefficient}
\begin{split}
    \ln{\phi} = Z - 1 - \ln{(Z - B)} -  \\
    - \frac{A}{2\sqrt{2} B} \ln{\left( \frac{Z + 2.414 B}{Z - 0.414 B} \right)}
\end{split}
\end{equation}

The fugacity coefficient $\phi$ represents the correction factor required to transform the ideal gas pressure $P$ to the real gas pressure, also called fugacity. At saturation, the fugacity of the liquid phase and that of the vapor should be equal to maintain equilibrium. Thus, it is possible to solve eq. \ref{eq:peng-robinson-polynomial} at any temperature $T$ for a pressure $P$ that results in equal fugacities for the liquid and vapor phases. 

\item{PRSV1 and PRSV2}

While the Peng-Robinson equation provides great results when used for vapor-liquid equilibrium calculations, its usage on some molecules results in deviations. For this reason, in 1986 two modifications to the original Peng-Robinson equation were proposed, known as PRSV1 (Peng-Robinson-Stryjek-Vera) and PRSV2,  \cite{Stryjek1986} \cite{Stryjek19862}, both of which changed eq. \ref{eq:alpha_preos} and added molecule-specific correction factors. These equations also made use of an improved variant of eq. \ref{eq:kappa}, given by:

\begin{equation} \label{eq:kappa_0}
\begin{split}
    \kappa_{0} = 0.378893 + 1.4897153 \omega - \\
    - 0.17131848 \omega^{2}+ 0.0196544 \omega^{3}
\end{split}
\end{equation}

In the case of the PRSV1 equation, the modification came in the form of a more complex $\kappa$ function, presented as:

\begin{equation} \label{eq:kappa_prsv1}
    \kappa = \kappa_{0} + \kappa_{1} \left( 1 + \sqrt{\frac{T}{T_{C}}} \right) \left(0.7 - \frac{T}{T_{C}} \right)
\end{equation}

\noindent where $\kappa_{1}$ is a molecule specific parameter, the source material providing a large list of already determined values. This form of the equation can be used universally for temperatures up to 70\% of the critical temperature. Above that point, molecules other than water and alcohols benefit from using $\kappa_{1} = 0$. 

Similarly, PRSV2 comes with a more complex $\kappa$ function, given by:

\begin{equation} \label{eq:kappa_prsv2}
\begin{split}
\kappa = \kappa_{0} + \left[ \kappa_{1} + \kappa_{2} \left( \kappa_{3} - \frac{T}{T_{C}} \right)\right. \cdot \\
 \cdot \left. \left( 1 - \sqrt{\frac{T}{T_{C}}} \right) \right] \left( 1 + \sqrt{\frac{T}{T_{C}}} \right) \left(0.7 - \frac{T}{T_{C}} \right) 
\end{split}
\end{equation}

\noindent with $\kappa_{1}$ the same as for PRSV1, and $\kappa_{2}$ and $\kappa_{3}$ molecule-specific parameters that are provided by the source material. 

As both PRSV1 and PRSV2 provide corrections to small deviations for vapor-liquid equilibrium calculations in the subcritical range, they can be used in the same manner as the Peng-Robinson equation above the critical point. It is thus expected that the small corrections will result in larger deviations far away from the critical point.

\item {The critical isochore}

Since the usage of gasses in their supercritical state in industrial applications is common, there is a constant demand for approximations of the curve of coexistence in the supercritical domain. However, due to the indistinguishability between vapors and liquids in the supercritical phase, artificial boundaries have to be determined to distinguish the two pseudo phases. For this reason, different authors investigated different approximation lines that were tied to already known functions and constants from thermodynamics, trying to relate them to liquid-like or vapor-like behavior \cite{Gorbaty1998}.

A popular choice became the isochore line that crossed the critical point, commonly referred to as the critical isochore. It has been shown that, in the case of water, the critical isochore corresponds to the phase diagram region where water molecules gained the possibility to rotate freely. While this state cannot be described as liquid-like, it is far away from gas-like behavior \cite{Gorbaty1998}.

While the author of the research mentions that the critical isochore serves as a good approximation for water only and advises against its usage in the case of typical liquids such as methane, the ambiguity of the problem leaves room for experimentation.

\item{The Widom-Banuti line}

Subsequent research on the continuation of the saturation pressure in the supercritical phase resulted in the concept of the Widom line, a line that divides liquid-like behavior from gas-like behavior. Based on numerical analysis of the scaling of the saturation pressure in the subcritical domain, Banuti produced an empirical formula for the Widom line \cite{Banuti2017}, given by:

\begin{equation} \label{eq:widom-banuti}
    P_{sat} = P_{C} \exp{ \left( \frac{A_{S}}{min\left(\frac{T}{T_{C}},1\right)} \left(\frac{T}{T_{C}} - 1\right) \right)}
\end{equation}

\noindent where $A_{S}$ is a species-dependent parameter that is related to the slope of the saturation pressure, with values provided by the source material. It is reported that the equation matches well with experimental measurements and can be accurate for pressures as high as three times the critical pressure.

\end{itemize}

\subsubsection{Adsorbate density}

Another challenge that comes with the potential theory is the determination of the adsorption volume, which is dependent on the density of the adsorbate in the adsorbed phase. As the vapor inside the nanoporous adsorbent is constrained to small volumes, it acts as a compressed vapor instead of a regular gas. For this reason, experimental data available on density cannot be used as a reference value as it does not display the expected behavior.

It has been proposed that the behavior of the density of the adsorbate is similar to that of a gas compressed at 100 MPa \cite{Seliverstova1986}, which should correspond to a nearly linear decrease of density with increasing temperature. 

\begin{itemize}

\item{Ozawa's method}

In 1976, while reviewing previous work done on adsorption theory, Ozawa proposed a density model, which should mimic that of a super-heated liquid \cite{Ozawa1976}, given by the following equation:

\begin{equation} \label{eq:ozawa}
    \rho = \rho_{b} \exp{( -0.0025 (T-T_{b}))}
\end{equation}

where $\rho_{b}$ is the density of the adsorbate at the boiling point, $T$ is the surrounding temperature, and $T_{b}$ is the boiling temperature. While the original formula included the thermal expansion coefficient $\alpha$ of the adsorbate, an average value of the thermal expansion of liquefied gases is used, which is determined to be $\alpha=2.5\cdot$10\textsuperscript{-3} K\textsuperscript{-1}. Since then, different methods have been proposed for the determination of the thermal expansion coefficient \cite{Srinivasan2011}, however, none of them provided any significant improvements in performance while still maintaining efficiency.

\begin{figure*}[!t]
    \centering
    \includegraphics[width=1\linewidth]{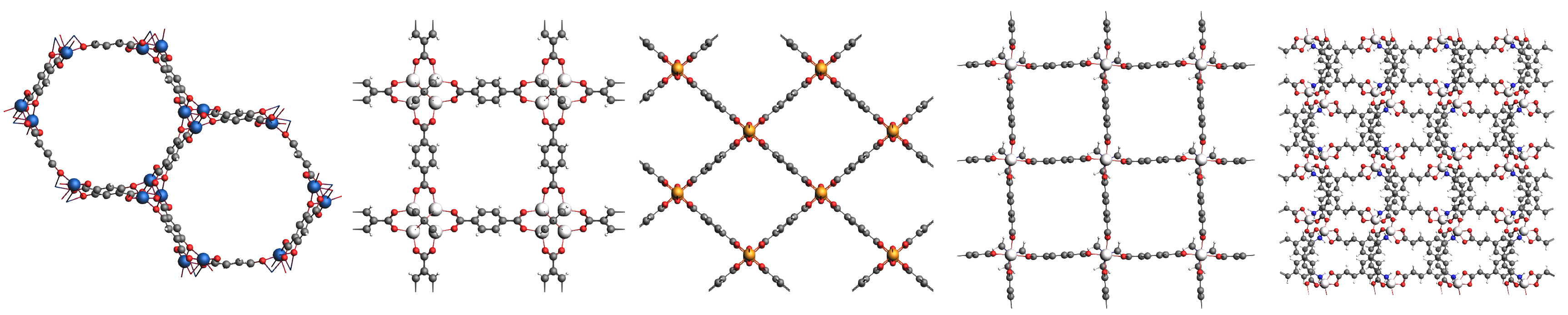}
    \caption{Snapshot of the MOFs used in the simulations. From left to right Co-MOF-74, IRMOF-1, MIL-47, MOF-1, and ZJU-198}
    \label{fig:mofs_structures}
\end{figure*}

\begin{table*}[!t]
    \centering
    \caption{Characterization of nanoporous materials used in the simulations; the large cavity diameter has been obtained from the pore size distribution of each MOF \cite{MOF1} \cite{MOF2} \cite{MOF3}.}
    \begin{tabular}{c c c c c c}
        {} & Co-MOF-74 & IRMOF-1 & MIL-47 & MOF-1 & ZJU-198 \\
        \hline
        \hline
        Pore volume (cm\textsuperscript{3}/g) & 0.598 & 1.373 & 0.610 & 0.762 & 0.267\\
        Framework density (kg/m\textsuperscript{3}) & 1180.6 & 593.4 & 1000.4 & 826.5 & 1247.3 \\
        Large cavity diameter (\AA) & 12 & 15 & 7 & 8.5 & 4 \\
        Helium void fraction (\%) & 70.6 & 81.5 & 61.0 & 62.9 & 33.3 \\
    \end{tabular}
    \label{tab:mofs}
\end{table*}

\item{Hauer's method}

Similar to the equation proposed by Ozawa, Hauer proposed in 2010 a linear equation that represented the linear decrease of the density \cite{Hauer2010}, given by:

\begin{equation} \label{eq:hauer}
    \rho_{ads} = \rho_{r} (1 - \alpha (T-T_{r}))
\end{equation}

where $T_{r}$ is the reference temperature, $\rho_{r}$ is the density at the reference temperature, and $\alpha$ the thermal expansion coefficient. Unfortunately, the source material does not provide a systematic methodology for determining the reference temperature or the expansion coefficient, the reason for which alternative methods have to be developed. Coincidentally, the formula above corresponds to the one Dubinin proposed for an adsorbate in the adsorbed phase, although that equation uses a different coefficient from the thermal expansion coefficient and replaces the reference temperature with the boiling one \cite{Dubinin1960}.  Following Ozawa's and Dubinin's methods ideas, we set the $T_{r}$ and the $\rho_{r}$ as the temperature and density of the adsorbates in the boiling point. The selection of the value of the thermal expansion coefficient $\alpha$, will be discussed in the Results section.

\item{Empirical method}

While studying the adsorption of methane in high-rank coal, an empirical density model was proposed, which remains constant with temperature \cite{Meng2016}, given by:

\begin{equation} \label{eq:empirical}
    \rho_{ads} = \frac{8 P_{C}}{R T_{C}} M
\end{equation}

where $M$ is the molecular mass of the adsorbate, and $R$ is the universal gas constant.

\end{itemize}

\subsection{Simulation Details}

As the saturation pressure is determined below the critical point, molecules with lower critical temperatures are more prone to deviations in the virtual saturation pressure. This, in turn, results in a higher divergence of the characteristic curve. For this reason, molecules that have a low critical temperature can be used to test the validity of each model far away in the supercritical phase, while others with higher critical temperatures can be used as a reference. As such, three adsorbates that have industrial applications and varied critical temperatures are chosen for the analysis: carbon dioxide, methane, and molecular nitrogen. Apart from these, adsorbates such as argon, propane, ammonia, and methanol are also investigated; however, the respective isotherms are taken from the literature.

We select five MOFs, namely Co-MOF-74, IRMOF-1, MIL-47, MOF-1, and ZJU-198. These MOFs cover a range of shapes, topologies, chemical compositions, the presence of active sites of adsorption, pore sizes, and pore connectivities, properties that could influence the packing of the confined molecules. This packing would affect the density of the molecules in the adsorbed state, which is an important part of Polanyi's model. Co-MOF-74 and MIL-47 show a one-dimensional pore network composed of independent longitudinal channels. In terms of their size and properties, Co-MOF-74 has almost double the pore size of MIL-47 and also contains a high concentration of open metal sites. IRMOF-1 and MOF-1 show similar pore networks of interconnected cavities linked by windows. While they share the same chemical composition, Co-MOF-74 and MIL-47 differ in terms of pore sizes. Finally, ZJU-198 is a complex interpenetrated structure that has small cavities, which are similar to the size of the gases under study. An illustration of each MOF is presented in Figure \ref{fig:mofs_structures}, and their physical characteristics can be found in Table \ref{tab:mofs}. To complement the study and expand the capabilities of the proposed method, other structures taken from experimental samples are also investigated, such as MOF-467, Cu-BTC, LTA4A, Co$_3$-HCOO$_6$, Cu-INAIP, and STAM-1.

We used the molecular simulation software RASPA \cite{Sofia2016} to compute the adsorption isotherms for different temperatures, adsorbents, and adsorbates. The GCMC simulations run 50000 simulation cycles after a short initialization of 5000 cycles. For the small molecules under study, this amount of cycles ensures reasonable equilibration of the results, with the adsorbed amount fluctuating around an average value that results in a smooth behavior of the computed isotherms. Moreover, the structures used in this project are considered rigid. The interaction energies between molecules and adsorbates are described by Lennard-Jones and Coulombic potentials. The Lennard-Jones parameters of the adsorbents are taken from a combination of UFF (metal centers) \cite{UFF} and DREIDING (organic linkers) \cite{DREIDING} force-fields. The partial charges of each adsorbent have been reported in the literature \cite{MOF1} \cite{MOF2} \cite{MOF3}. Similarly, the molecular models for carbon dioxide \cite{carbon_dioxide}, methane \cite{methane}, and nitrogen \cite{nitrogen} have also been taken from previous works.

After obtaining the adsorption isotherms at 273, 300, 333, and 373 K, we converted them into adsorption characteristic curves using equations \ref{eq:potential} and \ref{eq:volume}. Starting with the characteristic curve obtained from the adsorption isotherms at 300K for each adsorbent/adsorbate pair, we reversed the calculation to reconstruct the adsorption isotherms at the other specified temperatures. These newly predicted isotherms were then compared to the reference results obtained from the GCMC simulations. After benchmarking all the methods that describe the saturation pressure and density defined above, we repeated the process using experimental results taken from the literature, employing the most promising approach.

To assess the effectiveness of each combination of methods, we conducted an error analysis on the characteristic curves. We plotted the adsorption volume of the reference characteristic curve against the other curves and measured their linearity using two metrics: the correlation coefficient, denoted as $r$, and the residual sum of squares, RSS. The correlation coefficient $r$ can take values between -1 and 1, where 1 indicates a perfect correlation between the reference and predicted data points. Given that, in most cases, the characteristic curves should follow similar trends, a high correlation coefficient approaching 0.99 was expected for nearly all cases, demanding the use of four significant digits for an adequate comparison. Regarding the residual sum of squares, lower values indicate fewer deviations from the reference line, while higher values are associated with greater deviations and, consequently, poorer predictions.

\section{Results and Discussion}
\label{sec:results}

The initial step in applying the adsorption potential theory involves developing models to describe both the vapor saturation pressure and the density of molecules in the adsorbed phase. These models should be able to accurately depict these properties across the desired range of operating conditions, including conditions above and below the critical temperature and pressure. In this study, we set a reference temperature under ambient conditions to predict adsorption data, ranging from 273 K to 373 K, which is practical for various applications. It is important to note that this temperature selection does not restrict the applicability of the approach proposed in this study to other ranges of working conditions; however, the saturation pressure and density need to be defined within the chosen temperature interval.

\begin{figure}[!t]
    \centering
    \includegraphics[width=0.48\textwidth]{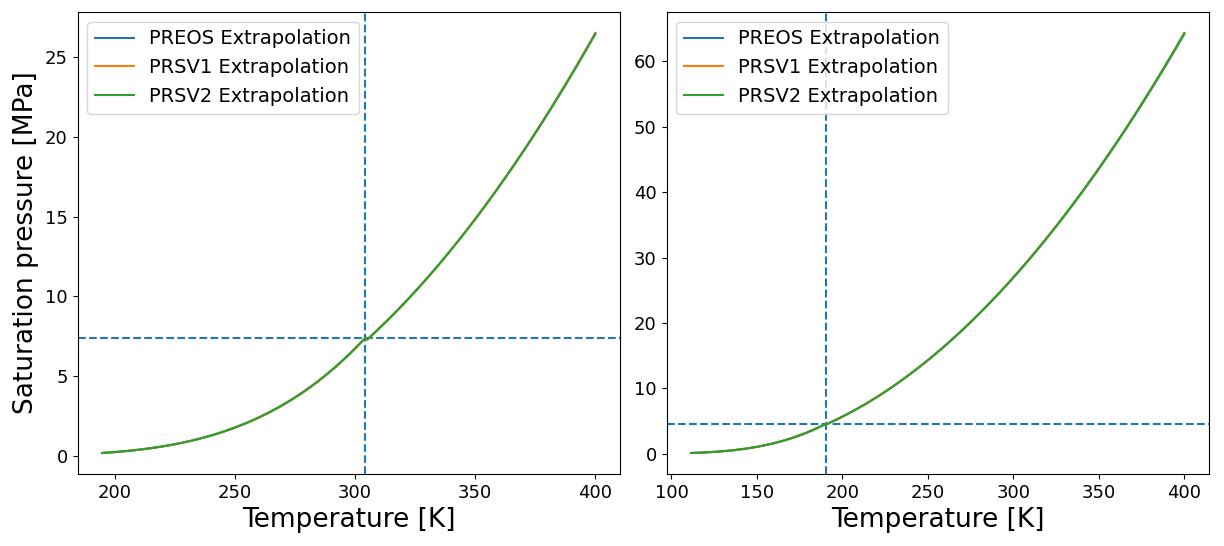}\\
    \caption{Comparison between the PREoS, PRSV1, and PRSV2 for carbon dioxide (left) and methane (right) using a second order polynomial for extrapolation. The vertical dotted line represents the critical temperature, and the horizontal dotted line represents the critical pressure.}
    \label{fig:Fig_02}
\end{figure}

Regarding the saturation pressure, our investigation initially focuses on the Peng Robinson equation of State (PREoS) and its derivatives PRSV1 and PRSV2. While the PREoS is primarily designed for determining the saturation pressure in the subcritical range, there are no apparent limitations to its applicability above this range, referred to as the virtual saturation pressure. However, we observed an exponential increase in saturation pressure values when solving for conditions above the critical point, eventually leading to numerical divergence. Consequently, we used an extrapolation method based on a second-order polynomial as a reliable approach when using the PREoS (along with PRSV1 and PRSV2) above the critical point. Figure \ref{fig:Fig_02} shows the vapor-liquid equilibrium line for carbon dioxide and methane, starting from their respective boiling points up to 400 K. Notably, the PREoS, PRSV1, and PRSV2 predictions align closely on the same saturation pressure curve. The deviations introduced by the additional parameters in the PRSV1 and PRSV2 equations exhibit minimal differences from the original PREoS. For instance, in the case of methane at 400 K, more than double its critical temperature (191 K), both PRSV1 and PRSV2 remain within 0.2\% of the PREoS values. Given the necessity for empirical constants that are undefined for many molecules, the impracticality of PRSV1 and PRSV2 methods becomes evident, leading to their exclusion from the rest of the analysis.

\begin{figure}[!t]
    \centering
    \includegraphics[width=0.48\textwidth]{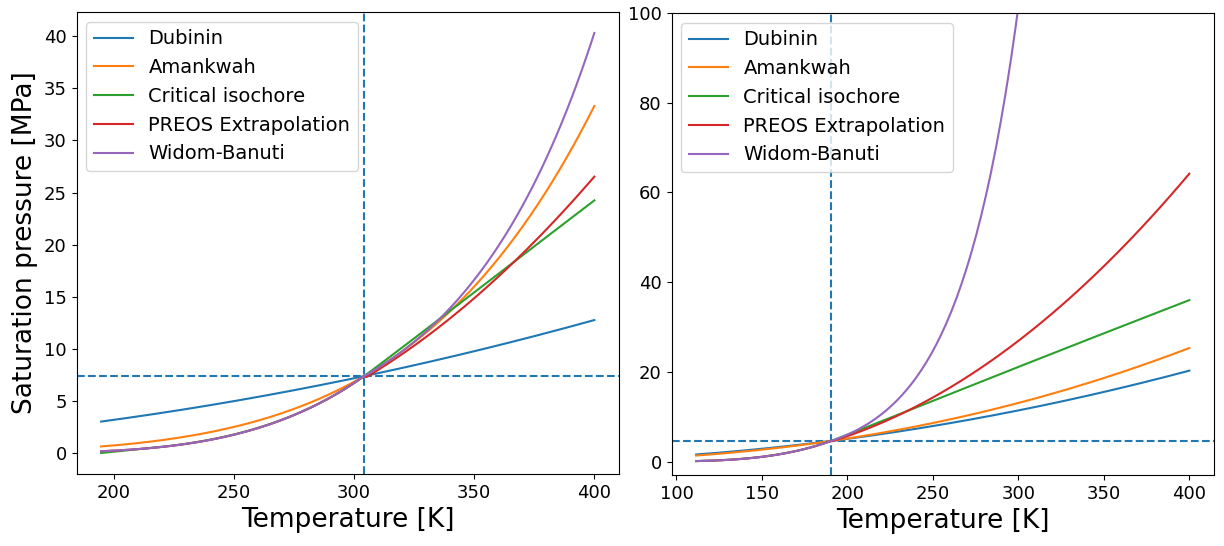}\\
    \caption{Comparison between Dubinin's method, Amankwah's method, the critical isochore, extrapolation of the coexistence curve obtained using the PREoS using a second order polynomial for extrapolation, and the Widom-Banuti line for carbon dioxide (left) and methane (right). The vertical dotted line represents the critical temperature, and the horizontal dotted line represents the critical pressure.}
    \label{fig:Fig_03}
\end{figure}

\begin{figure}[!t]
    \centering
    \includegraphics[width=0.30\textwidth]{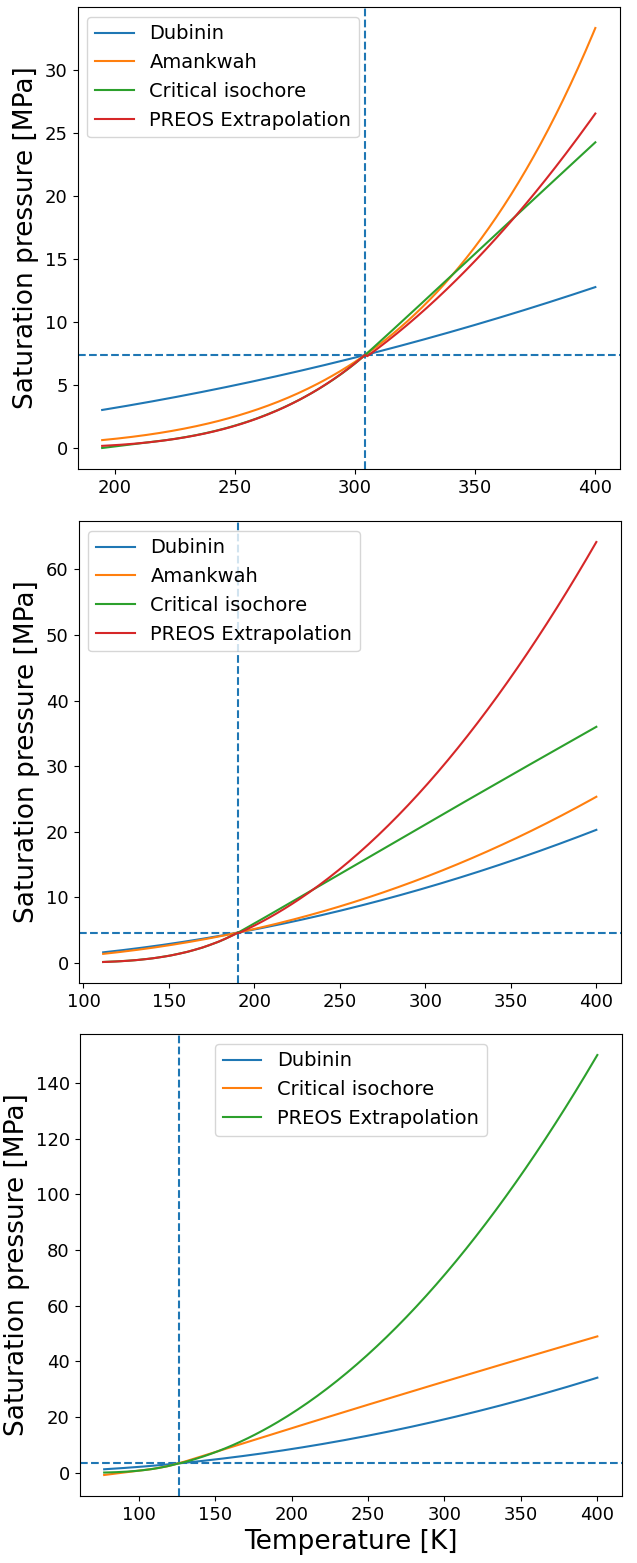}\\
    \caption{Comparison between Dubinin's method, Amankwah's method, the critical isochore, extrapolation of the coexistence curve obtained using the PREoS using a second order polynomial for extrapolation for carbon dioxide (top) and methane (center), and nitrogen (bottom). The vertical dotted line represents the critical temperature, and the horizontal dotted line represents the critical pressure.}
    \label{fig:Fig_04}
\end{figure}

With the removal of the PREoS derivatives, we proceed to examine the saturation pressure using the remaining methods: Dubinin's method, Amankwah's method, the critical isochore, extrapolation of the coexistence curve derived from the Peng-Robinson method, and the Widom-Banuti line. While all these methods are expected to diverge above the critical point, they should still maintain a similar order of magnitude (Figure \ref{fig:Fig_03}). For the sake of analysis, Amankwah's method serves as a reference, given that its coefficients were initially fitted to experimental data, ensuring a fair estimation of the saturation pressure \cite{Song2018}. In a preliminary assessment, we observe that all curves exhibit similar behavior, except for the Widom-Banuti line for methane, which shows substantial deviations two orders of magnitude greater than the others, revealing an exponential rise with temperature. These findings also shed light on the physical interpretation of the virtual saturation pressure within the context of the adsorption potential theory. The Widom-Banuti line should accurately represent the boundary between liquid-like and gas-like behavior at temperatures below three times the critical temperature. This boundary may hold less relevance for the adsorption potential theory, which primarily concerns behavior between gas and non-gas phases. The critical isochore better captures this situation, consistent with the accuracy observed in predictions of adsorption data as described below. As a result, we exclude the Widom-Banuti line from further consideration in the study.

Figure \ref{fig:Fig_04} shows a comparison of the remaining methods, with nitrogen included as a test adsorbate due to its relatively low critical temperature (126 K) compared to the other adsorbates. For carbon dioxide, all methods, except for Dubinin's model, exhibit a similar trend below and above the critical point. As anticipated, divergence becomes evident above the critical point, with values at 400 K deviating by approximately 5 MPa. This deviation is more pronounced for molecules with lower critical temperatures, with deviations at 400 K reaching 30 MPa for methane and 100 MPa for nitrogen. The primary source of these deviations arises from the extrapolation of PREoS results, suggesting that a second-order polynomial may lead to a too abrupt increase far from the critical point. Consequently, alternative functions could be explored. Nevertheless, for the purposes of this study, the second-order polynomial is considered as an option. Ideally, below the critical temperature, all the methods should converge to the experimental saturation pressure. However, this is only true for the critical isochore and the PREoS. On the contrary, Amankwah's and Dubinin's methods exhibit noticeable deviations in the experimentally affordable region, i.e., below the critical point, thus revealing their limitations. 

The second property of interest is the density of adsorbates when confined within the pores of the adsorbents. Due to the proximity in size between the cavities and the adsorbed molecules, gases adsorbed within these spaces may exhibit unexpected behavior as temperature or pressure increases. Consequently, conventional models or equations of state are inadequate for modeling the density of the adsorbate in such scenarios. Fortunately, the adsorption theory offers a workaround to assess whether changes in density align with the adsorbate-adsorbent systems. This workaround is conceived in the concept that the volume filling of the adsorbent relies on the energy required to condense the fluids. If the environmental conditions are sufficient to induce condensation, the entire structure should be filled with adsorbate. Consequently, at saturation conditions, the adsorption volume should remain constant. Therefore, when measuring adsorption at saturation conditions, the relative change in loading between temperatures should closely mirror the relative change in the density of the adsorbate. To explore this phenomenon, we conducted Grand Canonical Monte Carlo (GCMC) simulations to obtain adsorption data at saturated conditions for carbon dioxide, nitrogen, and methane within the five selected MOFs across temperatures ranging from 253 K to 393 K.

\begin{figure}[!t]
    \centering
    \includegraphics[width=0.48\textwidth]{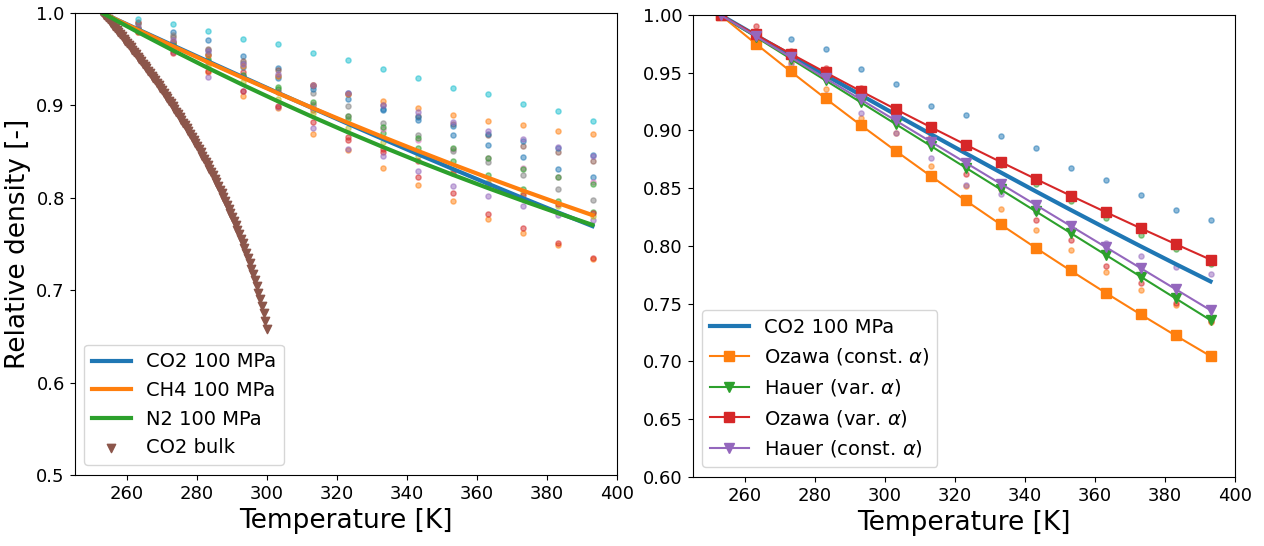}\\
    \caption{Comparison between the relative density of carbon dioxide, methane, and nitrogen in the five MOFs (left) and in bulk under high pressure (100 MPa). The density of carbon dioxide in the bulk at saturation conditions is included for comparison. The figure on the right includes the relative density of carbon dioxide in the five MOFs and in the bulk at 100 MPa, together with the two variations of the density models of Hauer and Ozawa.}
    \label{fig:Fig_05}
\end{figure}

Figure \ref{fig:Fig_05} depicts the relative changes in loading at saturation pressures as temperature increases. Upon normalization at a specified temperature, these values should resemble the behavior of the density of adsorbates within the pores. We chose 253 K as the reference temperature for normalization since it covers the entire range of selected working conditions. It is worth noting that obtaining the real density value from these simulations is challenging, as it requires knowledge of the available pore volume for each molecule within each adsorbent. This calculation is non-trivial, as pore volume estimation is typically available only for a few molecules that exhibit low interactions with the adsorbents, such as helium, argon, or nitrogen. Given the complexity of the pore network in nanoporous materials, some molecules may be adsorbed in cavities or spaces inaccessible to others. Therefore, the term "effective" or "apparent" density is more suitable for the density computed in this work. Nevertheless, this is not considered a limitation in this context, as we employ the theory as a predictive tool to convert isotherms from one temperature to another. If there is a need for determining the working volume (see eq. \ref{eq:volume}), a more comprehensive analysis of the density in the adsorbed phase would be required.

Figure \ref{fig:Fig_05} (left) shows the relative density, i.e., the loading normalized to the maximum loading observed at 253 K, for carbon dioxide, methane, and nitrogen within the five selected MOFs. For comparison, we include the density of carbon dioxide in the bulk at saturation conditions, which is unsuitable for usage within the adsorption potential formulation. Seliverstova et al. \cite{Seliverstova1986} found that adsorbates confined within nanopores are compressed like a liquid under high pressure, such as 100 MPa. They also note that the average density of adsorbates in micropores depends intricately on the value of the adsorption field, which induces additional compression to the fluids. This compression results in a linear decrease in density as temperature increases within the range studied here. Consequently, we also include the experimental density of the three fluids at 100 MPa in Figure \ref{fig:Fig_05} (left), which is normalized to 253 K. It's evident that the relative decrease in loading with temperature aligns with the high-pressure fluid model when molecules are confined within the pores of MOFs with diverse pore sizes and shapes. However, slight variations in slope are noticeable among different adsorbate-adsorbent pairs. The lowest slope is observed for carbon dioxide in MOF-1 at -2.0$\cdot$10\textsuperscript{-3} K\textsuperscript{-1}, while the upper limit is for methane in ZJU-198 at -0.8$\cdot$10\textsuperscript{-3} K\textsuperscript{-1}. This implies that interactions between adsorbates and adsorbents are unique and, in principle, would require distinct density models. It is possible to compute the thermal expansion coefficients in eq. \ref{eq:ozawa} and eq. \ref{eq:hauer} using the decrease in loading for each adsorbate-adsorbent pair. However, this study is primarily focused on providing a transferable working model for predicting adsorption data, then, employing individual coefficients for each pair would be impractical. 

\begin{figure}[!t]
    \centering
    \includegraphics[width=0.48\textwidth]{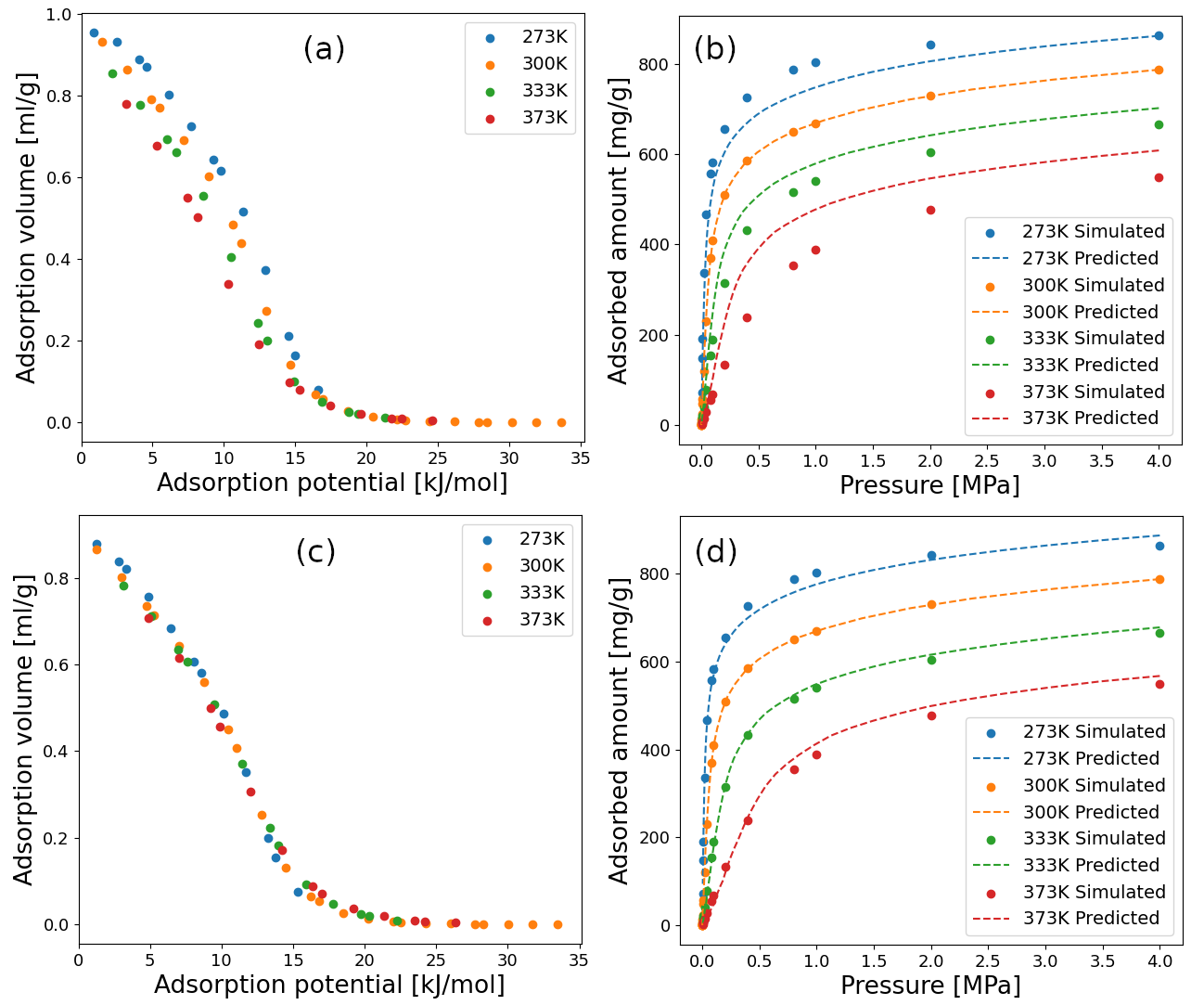}\\
    \caption{Characteristic curves (a,c) and adsorption isotherms (b,d) of carbon dioxide in MOF-1, using Dubinin’s and Ozawa’s methods with a constant thermal expansion coefficient, for saturation pressure and density, respectively (a,b), and the critical isochore and Hauer’s method using a universal thermal expansion coefficient, for saturation pressure and density, respectively (c,d). The predicted adsorption isotherms were reconstructed from the characteristic curve at 300 K.}
    \label{fig:Fig_06}
\end{figure}

Our strategy is to use an average slope value for each adsorbate, which will substantially simplify the method. The average thermal expansion coefficient values are 1.7$\cdot$10\textsuperscript{-3} K\textsuperscript{-1} for carbon dioxide, 1.3$\cdot$10\textsuperscript{-3} K\textsuperscript{-1} for methane, and 1.2$\cdot$10\textsuperscript{-3} K\textsuperscript{-1} for nitrogen. Given that all three gases exhibit the same decrease in density at high pressures, a universal thermal expansion coefficient of 1.65$\cdot$10\textsuperscript{-3} K\textsuperscript{-1} for all species is also proposed for the analysis. We carried out a parametric optimization process of $\alpha$ in equation \ref{eq:hauer} to derive this universal value by minimizying the deviations across all cases studied. To do so, we iteratively use the approach described in the Methodology to examine all combinations of adsorbents and adsorbates while varying $\alpha$. Subsequently, the results were ranked based on the correlation coefficient and residual sum of squares, which determined a universal thermal expansion coefficient of 1.65$\cdot$10\textsuperscript{-3} K\textsuperscript{-1} as the best-performing value. As a result, for the remainder of the study, four temperature density models are utilized: Ozawa's method with a constant thermal expansion coefficient of 2.5$\cdot$10\textsuperscript{-3} K\textsuperscript{-1}, or a custom expansion coefficient for each adsorbate, and Hauer's method with a universal thermal expansion coefficient of 1.65$\cdot$10\textsuperscript{-3} K\textsuperscript{-1}, or a custom expansion coefficient for each adsorbate. The scaling of these methods for carbon dioxide is depicted in Figure \ref{fig:Fig_05} (right), where the models remain close to the estimated density of carbon dioxide within the pores of the five MOFs.

With all the components of the adsorption potential theory in place, we employ the approach detailed in the Methodology to predict isotherms under different conditions. When predicting isotherms, the degree of overlap between the characteristic curves of each isotherm is closely linked to the accuracy of the predictions. This overlap is achievable only when employing an appropriate combination of saturation pressure and density models. Figure \ref{fig:Fig_06} illustrates the characteristic curves and a comparison between the simulated and their respective predicted isotherms for carbon dioxide in MOF-1 using two distinct variants. The first variant utilizes Dubinin's method for saturation pressure and Ozawa's method with a constant thermal expansion coefficient for density. The evident divergence in the characteristic curve is mirrored in the poor quality of the predicted isotherms. In general, deviations at low values of the adsorption potential impact the prediction of the maximum loading achieved, while deviations at low adsorption volume levels lead to inaccurate predictions at low pressures. The second variant applies the critical isochore method for saturation pressure and Hauer's model with a constant thermal expansion coefficient for density. In this case, all characteristic curves converge into a single curve, translating into reasonably accurate predictions of the isotherms with minor deviations. It is important to note that these predictions can be further refined by adjusting the density model for each specific adsorbent/adsorbate pair. However, such customization would compromise the method's transferability and, consequently, its potential applicability to a wide range of scenarios.

\begin{figure}[!t]
    \centering
    \includegraphics[width=0.35\textwidth]{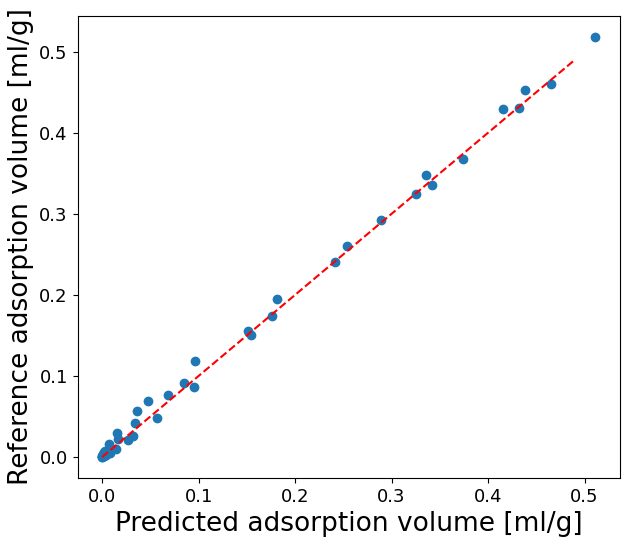}\\
    \caption{Reference against predicted adsorption volume at the same adsorption potential for nitrogen in MIL-47 using the characteristic curve at 300K as reference.}
    \label{fig:Fig_07}
\end{figure}

We have used the proposed approach to perform isotherm predictions for carbon dioxide, methane, and nitrogen in the five selected MOFs, using twenty density and saturation pressure model combinations. As previously emphasized, the degree of overlap in the characteristic curve is a crucial factor impacting prediction performance. To assess this overlap, we quantified it by calculating the correlation coefficient ($r$) and the residual sum of squares (RSS), as outlined in the Methodology. Figure \ref{fig:Fig_07} provides a representative illustration of the linearity in the assessment of adsorption volume. Our findings consistently indicate that the combination of the critical isochore for saturation pressure description, paired with Hauer's method utilizing a universal thermal expansion coefficient for density, surpasses the performance of all other possible combinations. The averaged correlation coefficient between carbon dioxide, methane, and nitrogen across all MOFs stands at $r$ = 0.9980, accompanied by a residual sum of squares of RSS = 0.0162.

The results detailed above strongly support the utilization of the proposed approach, based on the critical isochore and Hauer's method incorporating a universal thermal expansion coefficient for saturation pressure and density, respectively. Importantly, this approach is independent of the specific adsorbents and adsorbates, emphasizing the desired transferability of a predictive framework. Consequently, for the remainder of this study, we concentrate on implementing these models for saturation pressure and density. Amankwah's method and the PREoS present alternative options for estimating the virtual saturation pressure. We do not recommend Amankwah's method due to the requirement to define the exponential parameter for each specific adsorbate. This limitation lowers the transferability and predictive capabilities of the approach. As for PREoS, we will discuss its applicability at the end of the results section in this paper. At the same time, Ozawa's method stands as an alternative to Hauer's density model, often yielding similar performance. Nevertheless, several factors favor Hauer's model. Firstly, the linear formulation of Hauer's model is simpler compared to the exponential function found in Ozawa's model. Secondly, Hauer's method more effectively captures the observed linear decrease in density with increasing temperature, as documented in experiments.\cite{Seliverstova1986} Ultimately, due to the unsatisfactory performance of the combination involving Dubinin's method for saturation pressure and an empirical method for density, we opt to exclude these models from the scope of this work.

\begin{figure}[!t]
    \centering
    \includegraphics[width=0.48\textwidth]{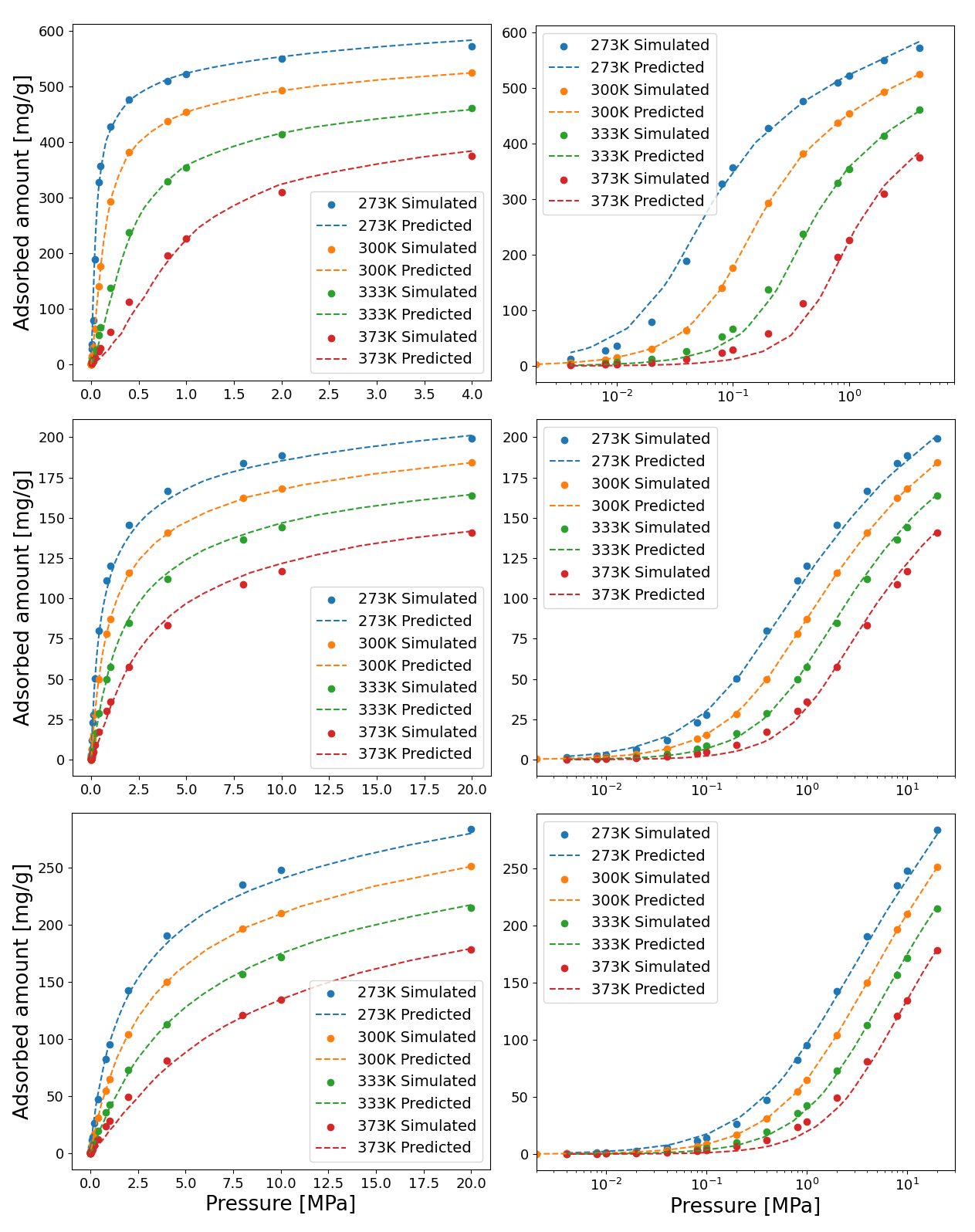}\\
    \caption{Linear (left) and logarithmic (right) representation of the prediction of the adsorption isotherms for carbon dioxide (top), methane (center), and nitrogen (bottom) in MIL-47, using the critical isochore and Hauer’s method using a universal thermal expansion coefficient, for saturation pressure and density, respectively.}
    \label{fig:Fig_08}
\end{figure}

In Figure \ref{fig:Fig_08}, we show the predictive capabilities of this approach for the adsorption of carbon dioxide, methane, and nitrogen in MIL-47. The figure illustrates the results in linear and logarithmic representations, emphasizing the excellent agreement between the simulated and predicted isotherms across the entire pressure range. While the linear graph provides deeper insights into the quality of predictions at high pressures, it may not depict low-pressure areas with the same clarity. In contrast, the logarithmic representations offer a comprehensive view of the predictions at the expense of scale legibility.

We have used the isotherms of each molecule at 300 K as the reference isotherms for predicting isotherms at other temperatures. It is essential to note that the uniqueness of the characteristic curve allows for similar performance using different temperatures as references. The same principle extends to the range of operating conditions. While we chose a temperature interval between 273 and 373 K as a study case, the primary restriction relies on the range of pressure and temperature values within which we can accurately model the saturation pressure of the molecules. However, we advise employing the framework at conditions near the selected reference temperature, which naturally produces more accurate predictions. Any deviation between the slope of the density model and the actual density becomes more significant as we move further from the reference temperature. It is crucial to emphasize that, to mitigate these deviations at temperatures distant from the reference, this approach allows for fine-tuning the model for each specific adsorbent/adsorbate pair. This room for improvement can prove invaluable in situations demanding high accuracy at the expense of some transferability to other working pairs. In Figure \ref{fig:Fig_08}, there is no distinction in the framework between the three adsorbates, apart from considering their individual physicochemical properties, which, in this instance, are reduced to the critical isochore, the boiling temperature, and its corresponding density needed for eq. \ref{eq:hauer}. This highlights the transferability of the approach between different adsorbates. Furthermore, we have found similar performance for the adsorption of these three molecules in the other MOFs under study, thus expanding the mentioned transferability to adsorbents characterized by varying pore sizes, shapes, and chemical compositions.

\begin{figure*}[!t]
    \centering
    \includegraphics[width=0.75\textwidth]{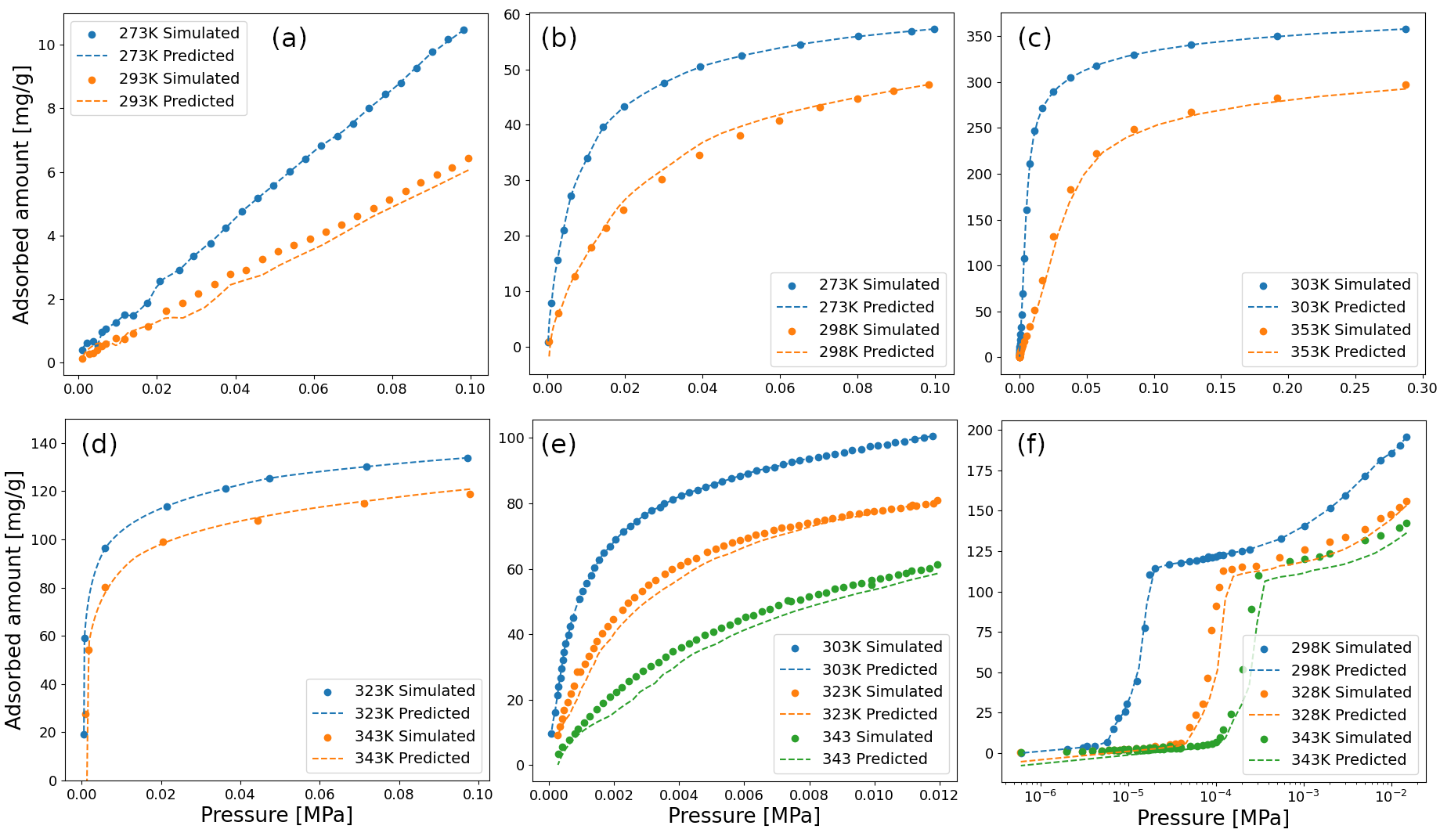}\\
    \caption{Performance of the method for predicting adsorption isotherms of argon in MOF-467 \cite{argon_in_MOF-467}, (a), carbon dioxide in Cu-INAIP \cite{CO2_in_Cu-INAIP}, (b), propane in Cu-BTC \cite{propane_in_Cu-BTC}, (c), ammonia in LTA4A \cite{ammonia_in_LTA4A}, (d), methanol in Co$_3$-HCOO$_6$ \cite{methanol_in_Co3}, (e), and methanol in STAM-1 \cite{methanol_in_STAM-1}, (f), using the critical isochore and Hauer’s method using a universal thermal expansion coefficient, for saturation pressure and density, respectively.}
    \label{fig:Fig_09}
\end{figure*}

The previous findings highlight the robustness of the approach presented in this work for predicting adsorption isotherms through molecular simulations across a broad spectrum of conditions within MOFs, specifically for carbon dioxide, methane, and nitrogen. To further validate this method and assess its performance in predicting adsorption isotherms for diverse adsorbent/adsorbate pairs, we extended our analysis to experimental data obtained from the literature. We carefully selected polar and non-polar molecules with varying molecular sizes and critical temperatures, as these properties significantly influence the modeling of density and saturation pressure. The chosen molecules for this validation included argon, propane, ammonia, and methanol.

Figure \ref{fig:Fig_09} compares the experimental and predicted adsorption isotherms for these molecules in various porous materials. In all instances, we utilized the isotherms at lower temperatures as reference data to compute the corresponding isotherms at higher temperatures. It is important to note that the experimental values obtained from the literature (taken from the NIST adsorption database), may exhibit some degree of noise. This noise is propagated to our predictions as well. To mitigate this effect, one could consider fitting the measured data points to an appropriate isotherm equation to smooth out any tendencies in the data, thereby reducing noise-induced errors. Nevertheless, for the sake of a fair comparison between our approach and the available experimental data, we opted to use the reported data as-is. Despite the presence of such noise, it's evident that all the experimental and predicted curves are in excellent agreement.

Different mechanisms govern the adsorption behavior of polar and non-polar molecules. Non-polar molecules typically exhibit weak interactions both among themselves and with the adsorbents. For instance, we conducted a comparison of the adsorption isotherms of argon in MOF-467 \cite{argon_in_MOF-467}, carbon dioxide in Cu-INAIP \cite{CO2_in_Cu-INAIP}, and propane in Cu-BTC \cite{propane_in_Cu-BTC}, as depicted in Figures \ref{fig:Fig_09} (a, b, and c). In contrast, polar adsorbates often exhibit strong interactions between themselves and with the adsorbents. Notable examples include ammonia and methanol, molecules capable of forming hydrogen bonds during adsorption.\cite{MATITOMARTOS2020,flexible_structure,CS1000a-alcohols,MAF-6-alcohols} Figures \ref{fig:Fig_09} (d and e) highlight the successful prediction of adsorption isotherms for ammonia in LTA4A \cite{ammonia_in_LTA4A} and methanol in Co$_3$-HCOO$_6$ \cite{methanol_in_Co3}, reflecting the capability of our approach to model polar adsorbates with strong interactions accurately. Nevertheless, it is noteworthy that the nature of the adsorbents and adsorbates does not limit the applicability of this method. An interesting scenario is presented in Figure \ref{fig:Fig_09} (f), which shows the adsorption isotherms of methanol in STAM-1 \cite{methanol_in_STAM-1}. STAM-1 is a flexible MOF that undergoes a gate-opening mechanism when adsorbing polar molecules. This intricate process involves a structural transformation of the adsorbent induced by the adsorbed molecules, resulting in a sudden jump in loading at a specific pressure on the isotherm. Subsequently, the molecules gradually fill the cavities of the MOF.\cite{flexible_structure} Notably, our adsorption potential theory can effectively predict such complex mechanisms, including adsorption on flexible frameworks. However, this capability is restricted by the temperature's influence on the material's flexibility, requiring in-depth analysis in situations where temperature significantly impacts the framework's structural dynamics.

We have successfully demonstrated the transferability and accuracy of the adsorption potential theory based on our approach to modeling vapor saturation pressure and density in the adsorbed phase. However, it is important to realize that the main limitation of our method lies in finding the critical isochore line for a specific adsorbate. The critical isochore can be experimentally determined for many molecules, but for some, it may not be available. Consequently, we need an alternative method to model the saturation pressure if we want to apply this approach. Figures \ref{fig:Fig_03} and \ref{fig:Fig_04} illustrate that below the critical temperature, both the critical isochore and the (PREoS) yield nearly identical results that align with experimental vapor saturation pressure. This makes PREoS a potential alternative to the critical isochore, particularly because this equation is only based on the critical temperature, pressure, and acentric factors of the molecules. These properties are accessible to a large number of adsorbates. 

\begin{figure}[!t]
    \centering
    \includegraphics[width=0.35\textwidth]{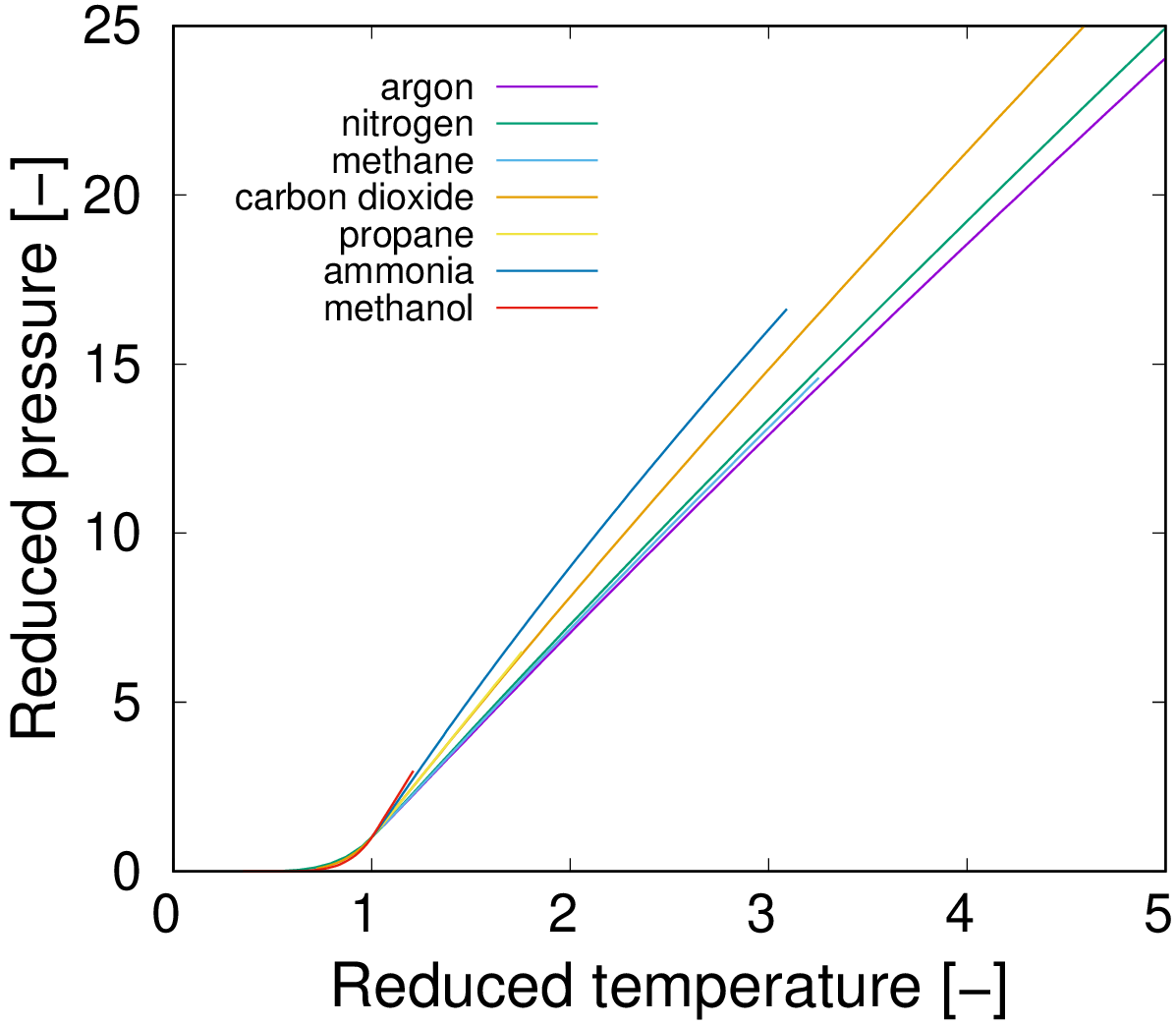}\\
    \caption{Reduced representation of the coexistence curve obtained using the critical isochore for the molecules studied in this work.}
    \label{fig:Fig_10}
\end{figure}

In our study, we utilized a second-order extrapolation as a representative means to model the virtual saturation pressure, i.e., values above the critical temperature of the adsorbates. However, we could approximate this interpolation using other functions that mimic the critical isochore of similar molecules. To shed light on this concept, Figure \ref{fig:Fig_10} displays the coexistence curve of the molecules examined here, normalized in terms of reduced pressure and temperature, i.e., relative to their respective critical pressure and temperature values. These curves reveal a linear relationship between the saturation pressure and temperature, characterized by similar slopes that gradually increase with the critical temperature of the molecule. We can distinguish three distinct trends: the lowest slopes correspond to nitrogen, argon, and methane with critical temperatures (T$_C$) of 126.2, 150.8, and 190.6 K, respectively. These are followed by carbon dioxide and propane with T$_C$ values of 304.1 and 369.9 K, respectively. The highest slopes are observed for ammonia and methanol with T$_C$ values of 405.5 and 512.6 K, respectively. In light of these findings, for molecules lacking an available critical isochore, we propose extrapolating the PREoS using a linear function that matches the slope of the reduced coexistence curve, as provided by the critical isochore of a similar molecule. This approach could potentially provide a useful solution to overcome the limitation associated with finding the critical isochore for specific adsorbates.

\section{Conclusions}
\label{sec:conclusions}

In this paper, we have investigated the adsorption properties of seven adsorbates and eleven porous materials to benchmark the performance of an adapted thermodynamical model for predicting adsorption equilibrium data. First, we employed Grand Canonical Monte Carlo simulations to compute adsorption isotherms for carbon dioxide, methane, and nitrogen in a few MOFs, namely Co-MOF-74, IRMOF-1, MIL-47, MOF-1, and ZJU-198. To complement the set of adsorption data and further validate our approach, we also take values from the literature for argon, propane, methanol, and ammonia in five additional MOFs (MOF-467, Cu-INAIP, Co$_3$-HCOO$_6$, STAM-1, and Cu-BTC) and one zeolite (LTA4A). We aimed to investigate Dubinin-Polanyi's adsorption model and propose enhanced approaches for modeling vapor saturation pressure and density within the cavities. To achieve this, we analyzed and assessed five different density models and four distinct vapor saturation pressure models for their potential usage in predicting adsorption isotherms. Based on our evaluation, we introduce a practical, refined thermodynamic model. This model involves utilizing the critical isochore as an approximation of the saturation pressure above the critical point, and applying Hauer's method with a universal thermal expansion coefficient (1.65$\cdot$10\textsuperscript{-3} K\textsuperscript{-1}) for the density of the adsorbed molecules. As an alternative to the critical isochore, we also propose an extrapolation procedure based on the Peng Robinson Equation of State to describe the required vapor saturation pressure. 

The validity of our method is demonstrated not only through simulated data but also through experimental data from the literature. This demonstrates its high transferability to various molecules and adsorbents. Moreover, our method consistently outperforms previously proposed methods in predicting adsorption equilibrium data, even in complex scenarios such as adsorption isotherms in flexible frameworks. Because the model couples pressure and temperature conditions in the adsorption potential definition, it is extendable to the prediction of adsorption isobars. The high accuracy of our predictions and the ability to retain the isotherm shape makes this approach suitable for industrial applications, particularly in cases involving high pressures. Furthermore, the model's simplicity, grounded in physically interpretable properties, allows minor adjustments to achieve improved results in specific circumstances.




\section*{Acknowledgements}

A.L-T and S.C. acknowledge funding from the Irène Curie Fellowship program of the Eindhoven University of Technology.



\section*{Competing interests}

The authors declare no competing interests.

\onecolumngrid

\bibliography{Isotherms_Prediction}













\end{document}